%% file: main.tex
\documentclass[useAMS,usenatbib]{mnras}

\input{include_tex/pacchetti}
\input{include_tex/journals}

\input{include_tex/definizioni}

\input{include_tex/additional_def}
\input{include_tex/colors}

\defcitealias{paramita:2018}{B18}
\defcitealias{DiMascia2021}{DM21}
\defcitealias{Valentini:2021}{V21}

\graphicspath{{figures/},{figures/SED_fitting/}, {figures/Td_PDF/}, {figures/scatter_plots/}}

\date{}
\pagerange{\pageref{firstpage}--\pageref{lastpage}} \pubyear{2021}

\title[Is the star formation rate in $z\sim 6$ quasars overestimated?]{Is the star formation rate in $z\sim 6$ quasars overestimated?}
\author[Fabio Di Mascia et al.]{Fabio Di Mascia$^{1}$\thanks{\href{mailto:fabio.dimascia@sns.it}{fabio.dimascia@sns.it}}, Stefano Carniani$^{1}$, Simona Gallerani$^{1}$, Fabio Vito$^{2}$,  
\newauthor{Andrea Pallottini$^{1}$, Andrea Ferrara$^{1}$, Milena Valentini$^{3,4,5}$}
\\
$^{1}$Scuola Normale Superiore, Piazza dei Cavalieri 7, I-56126 Pisa, Italy\\
$^{2}$INAF – Osservatorio di Astrofisica e Scienza dello Spazio di Bologna, Via Gobetti 93/3, I-40129 Bologna, Italy\\
$^{3}$ Universit{\"a}ts-Sternwarte, Fakult{\"a}t f{\"u}r Physik,  Ludwig-Maximilians Universit{\"a}t  M{\"u}nchen, Scheinerstr. 1, D-81679 M{\"u}nchen, Germany\\
$^{4}$ Excellence Cluster ORIGINS, Boltzmannstr. 2, D-85748 Garching, Germany\\
$^{5}$INAF - Osservatorio Astronomico di Trieste, via G.B. Tiepolo, 11 I-34143 Trieste, Italy
}

\begin{document}

\maketitle

\label{firstpage}

\begin{abstract}
The large total infrared (TIR) luminosities ($L_{\rm TIR} \gtrsim 10^{12}~\lsun$) observed in $z \sim 6$ quasars are generally converted into high star formation rates ($\SFR \gtrsim 10^2~\msunyr$) of their host galaxies. However, these estimates rely on the assumption that dust heating is dominated by stellar radiation, neglecting the contribution from the central Active Galactic Nuclei (AGN).
We test the validity of this assumption by combining cosmological hydrodynamic simulations with radiative transfer calculations.  
We find that, when AGN radiation is included in the simulations, the mass (luminosity)-weighted dust temperature in the host galaxies increases from $T\approx 50$~K ($T \approx 70$~K) to $T\approx 80$~K ($T\approx 200$~K), suggesting that AGN effectively heat the bulk of dust in the host galaxy.
We compute the AGN-host galaxy $\SFR$ from the synthetic spectral energy distribution by using standard $\SFR - L_{\rm TIR}$ relations, and compare the results with the \quotes{true} values in the simulations. We find that the $\SFR$ is overestimated by a factor of $\approx 3$ ($\gtrsim 10$) for AGN bolometric luminosities of $L_{\rm bol} \approx 10^{12}~\lsun$ ($\gtrsim 10^{13}~\lsun$), implying that the star formation rates of $z\sim 6$ quasars can be overestimated by over an order of magnitude.
\end{abstract}


\section{Introduction}

In the last two decades hundreds of quasars have been discovered within the first Gyr of the Universe \citep[e.g.][]{Yang2020ApJ, Wang2021ApJ}. These objects are powered by accreting supermassive black holes (SMBHs, $10^{8-10}~\msun$, e.g \citealt{Wu2015Natur}), shining as Active Galactic Nuclei (AGN). 
Observations in the local Universe reveal that black hole masses correlate with the host-galaxy properties \citep[e.g.][]{Magorrian1998AJ, Ferrarese2000ApJ, Gebhardt2000ApJ, Marconi2003ApJ, Gultekin2009ApJ}, suggesting a co-evolution between the the galaxy and the central SMBH \citep[e.g.][]{kormendy2013, Harrison2017NatAs}, possibly mediated by a form of feedback from the AGN via energy/momentum injections onto the surrounding gas \citep[e.g.][]{SilkRees1998, King2003ApJ}. In this picture, AGN activity might also regulate the star formation in the host galaxy \citep[e.g.][]{Carniani2016, Harrison2018NatAs}. 
A connection between the two quantities is also suggested by the similar shape of the black hole accretion density and star formation rate density across cosmic time \citep[e.g.][]{Aird2015MNRAS.451.1892A}.

In the latest years, ALMA and NOEMA observations have provided the opportunity to study the properties of the interstellar medium (ISM) in several $z\sim 6$ bright quasar-host galaxies via their \CII$158~\mum$ and CO emission, as well as the underlying dust continuum \citep[e.g.][]{Gallerani2017PASA, Decarli:2018, Venemans:2018, Carniani2019MNRAS, Venemans2020ApJ, Neeleman2021ApJ}. These studies revealed the presence of large molecular gas reservoirs ($M_{\rm gas} \gtrsim 10^{10}~\msun$), dust masses ($M_{\rm dust} \approx 10^{7-8}~\msun$), and ongoing star formation rates ($\SFR s $) as high as $\approx 10^{2-3}~\msunyr$ (e.g. \citealt{Decarli:2018, Venemans2020ApJ}) in the host galaxies.

These estimates rely on the assumption that the far-infrared (FIR) emission is mainly due to stellar light reprocessed by dust. The FIR photometry is generally fitted with a grey-body function \citep[e.g.][]{Carniani2019MNRAS}, from which the dust mass $M_{\rm dust}$ is determined, whereas the dust temperature $T_{\rm dust}$ is assumed to be in the range $40-60$~K. The total infrared (TIR) luminosity, $L_{\rm TIR}$, is then computed by integrating the grey-body function in the wavelength range $8-1000~\mum$, and a $\SFR-L_{\rm TIR}$ calibration \citep[e.g.][]{Kennicutt:2012} is used to infer the star formation rate in the host galaxy. 
This computation is very sensitive to the value of $T_{\rm dust}$, which is often set a priori because in most cases only one or a few FIR photometric data points are available \citep[e.g.][]{Venemans2020ApJ}.

Recently, \citet{Walter2022ApJ} obtained high angular resolution measurements of the {\CII \ } emission and underlying dust continuum in the $z=6.9$ quasar \mbox{J2348-3054}. From the fit of the Spectral Energy Distribution (SED) they derived $T_{\rm dust} =84.9^{+ 8.9}_{-10.5}$~K, and then inferred a $\SFR$ of $4700~\msunyr$. This estimate is a factor of $\approx 10$ higher than the one suggested by the \CII-SFR relation in \citet{Herrera-Camus2018ApJ}. They also found $T_{\rm dust} \geq 134$~K in the innermost region ($\approx 110$~pc), implying an extreme star formation rate density of $10^4~\msunyr$~pc$^{-2}$, which is only marginally consistent with the Eddington limit for star formation \citep{Thompson2005ApJ}. The authors argued that this result might also be explained by a contribution from the central AGN to the dust heating in the innermost $\approx 100$~pc. 

Numerical simulations constitute a complementary tool to investigate the dust properties in the ISM of galaxies, thanks to the possibility to support them with radiative transfer calculations \citep[e.g.][]{Behrens:2018}. 
Recently, \citet{McKinney2021ApJ} simulated a post-merger dust-enshrouded AGN, representative of submillimiter galaxies (SMGs) at $z\sim 2-3$. They find that AGN can dominate the dust heating on kpc-scales, boosting the IR emission at $\lambda \gtrsim 100~\mum$ (typically associated with dust-reprocessed stellar light) by up to a factor of $4$. As a consequence, standard FIR-based calibrations would over-estimate the star formation rates by a comparable factor.
This finding is also supported by empirical works suggesting that the AGN contribution to the IR emission increases as a function of AGN power in $z \lesssim 2.5$ sources, questioning the extreme star formation rates ($\gtrsim 1000~\msunyr$) inferred for the IR-brightest sources \citep{SymeonidisPage2021MNRAS, Symeonidis2022arXiv220511645S}.

In this work, we test the goodness of the assumption that dust heating is dominated by stellar radiation in high-$z$ quasar-hosts exploiting numerical simulations. In \citet{DiMascia2021} -- hereafter \citetalias{DiMascia2021} -- we studied the AGN contribution to the dust heating in $z \sim 6$ galaxies by post-processing the cosmological hydrodynamical simulations by \citet{paramita:2018} -- hereafter \citetalias{paramita:2018} --  with the radiative transfer code \code{SKIRT} \citep{Baes:2015, Camps2016}. We found that in normal star-forming galaxies the dust temperature distribution is consistent with typical values adopted ($40-50$~K) in the SED fitting of $z>5$ sources. However, in runs including AGN radiation the dust temperature increases to $T_{\rm dust} > 100$~K in the regions closest ($\approx 200$~pc) to the AGN, consistent with the finding by \citet{Walter2022ApJ}.
We expand our previous work by analyzing the synthetic SEDs predicted by our simulations to investigate whether the SFR inferred from the IR emission is overestimated when the AGN contribution is not accounted for.

\section{Numerical Methods} \label{Numerical_methods}

The numerical model adopted in this work is similar to \citetalias{DiMascia2021} and we briefly summarize it below.

\subsection{Hydrodynamical simulations} \label{hydro_sim}

We adopt the suites of cosmological hydrodynamic simulations studied in \citetalias{paramita:2018} and in \citet{Valentini:2021} -- hereafter \citetalias{Valentini:2021}. These simulations make use of a modified version of the Smooth Particle Hydrodynamics (SPH) N-body code \code{GADGET-3} \citep{Springel:2005} to follow the evolution of a $\sim 10^{12}~\msun$ dark matter (DM) halo in a zoom-in fashion from ${z=100}$ to ${z=6}$. A flat {$\Lambda$}CDM cosmology is assumed, with: ${\Omega_{\rm M,0}= 0.3089}$, ${\Omega_{\rm \Lambda,0}= 0.6911}$, ${\Omega_{\rm B,0}= 0.0486}$, ${H_0 = 67.74~\rm{km~s}^{-1}~{\rm Mpc}^{-1}}$  \citep{PlanckCollaboration2016}.
The two suites differ for the particle resolution and subgrid physics implemented, as follows.

\subsubsection{\citetalias{paramita:2018} simulations} \label{B18_sims}

In \citetalias{paramita:2018} a DM-halo of $M_{\rm h} = 4.4 \times 10^{12}~\msun$ at $z=6$ (virial radius $R_{200} = 73$~pkpc\footnote{Throughout this paper ckpc refers to \emph{comoving} kpc and pkpc to \emph{physical} distances in kpc. When not explicitly stated, we are referring to physical distances.}) is chosen inside a comoving volume of ${(500~{\rm cMpc})^3}$ for re-simulation in a zoom-in region of ${(5.21}~{\rm cMpc})^3$. The highest resolution DM and gas particles in the zoom-in region have a mass of ${m_{\rm DM} = 7.54 \times 10^6~\msun}$ and ${m_{\rm gas} = 1.41 \times 10^6~\msun}$, respectively. Gravitational forces are softened on a scale of $\epsilon =1~h^{-1}$~ckpc, which corresponds to $\approx 210$~pc at $z=6$.
The physics of the ISM is described by the multiphase model of \citet{Springel:2003}, assuming a density threshold of ${n_{\rm SF} = 0.13 \ \cc}$ for star formation and a \citet{Chabrier:2003} initial mass function (IMF) in the mass range ${0.1-100~\msun}$. The code accounts for radiative heating and cooling, for stellar winds, supernova feedback and metal enrichment. Stellar evolution and chemical enrichment are computed following \citet{Tornatore:2007}. 

BHs are included in the simulation by placing a ${M_{\rm BH} = 10^5 \ h^{-1} ~\msun}$ BH seed at the centre of a ${M_{\rm h} = 10^9 \ h^{-1} ~\msun}$ DM halo, if it does not host a BH already. BHs can grow either by gas accretion or via mergers with other BHs. The former process is modelled via the Bondi-Hoyle-Littleton scheme \citep{Hoyle:1939, Bondi:1944, Bondi:1952}, and the accretion rate is capped at the Eddington rate ${\dot{M}_{\rm Edd}}$. A fraction of the accreted rest-mass energy is radiated away with a bolometric luminosity:
\begin{equation} \label{eq:luminosity_bh}
    L_{\rm bol} = \epsilon_{\rm r} \dot{M}_{\rm BH} c^2,
\end{equation}
where $c$ is the speed of light and $\epsilon_{\rm r}=0.1$ is the radiative efficiency. AGN feedback from the accreting BHs is modelled by distributing a fraction ${\epsilon_{\rm f} = 0.05}$ of the energy irradiated by the BHs onto the surrounding gas in kinetic form.
In this work we consider the runs \AGNsphere{} and \AGNcone{} from \citetalias{paramita:2018}, in which AGN feedback is distributed in a spherical symmetry and in a bi-cone with an half-opening angle of ${45\degree}$, respectively.

\subsubsection{\citetalias{Valentini:2021} simulations} \label{V21_sims}

In \citetalias{Valentini:2021} a DM-halo of $M_{\rm h}=1.12 \times 10^{12}~\msun$ is chosen for the zoom-in simulation inside a comoving volume of $(148~{\rm cMpc})^3$. A zoom-in region of size $(5.25~{\rm cMpc})^3$ is chosen for re-simulation, with the highest resolution particles of the zoom-in simulation having a mass of $m_{\rm DM}=1.55\times 10^6~\msun$ and $m_{\rm gas}=2.89 \times 10^5~\msun$. The gravitational softening lengths employed are $\epsilon_{\rm DM}=0.72$~ckpc and $\epsilon_{\rm bar}=0.41$~ckpc for DM and baryon particles respectively, the latter corresponding to $\approx 60$~pc at $z=6$, i.e. a factor of $\approx 3$ lower than in \citetalias{paramita:2018}.
Instead of the multiphase model by \citet{Springel:2003}, the ISM is described by means of the MUlti Phase Particle Integrator (MUPPI) sub-resolution model \citep[e.g.][]{Murante2010, Valentini2020MNRAS}. It features metal cooling, thermal and kinetic stellar feedback, the presence of a UV background, and a model for chemical evolution, following \citet{Tornatore:2007}. In particular, star formation is implemented with a \HH-based prescription instead of a density-based cryterion, as in \citetalias{paramita:2018}.

A fraction of the accreted rest-mass energy is radiated away from the BHs according to eq. \ref{eq:luminosity_bh}, assuming a radiative efficiency of $\epsilon_{\rm r} = 0.03$. A fraction $\epsilon_{\rm f} = 10^{-4}$ of the radiated luminosity $L_{\rm bol}$ is thermally coupled to the gas surrounding the BHs, and it is isotropically distributed into the gas.
BHs with ${M_{\rm BH} = 10^5 h^{-1}~\msun}$ are seeded in a $10^9 h^{-1} ~\msun$ BH-less DM halo, and they grow by accretion and mergers. 
In this work, we make use of the fiducial run of the suite by \citetalias{Valentini:2021}, which we will refer to as \AGNthermal{}.

\subsection{Radiative transfer} \label{RT_sim}

For each snapshot of the hydrodynamical simulations we identify all the AGN with $L_{\rm bol} > 10^{10}~\lsun$ as per eq. \ref{eq:luminosity_bh}. We choose for the RT post-processing a number of snapshots sufficient to well sample the AGN luminosity range between $10^{10-14}~\lsun$. We select for each RT simulation a cubic region of $60$~kpc size ($\sim 50$\% of the virial radius), centered on the center of mass of the most-massive halo. We choose snapshots from different simulations in order to make our results not biased by the subgrid physics and AGN feedback prescription implemented\footnote{Different feedback prescriptions can significantly affect the gas/metals distribution and the star formation, as already shown in the original work by \citetalias{paramita:2018} and \citetalias{Valentini:2021}, and also in \citetalias{DiMascia2021} (see their Figure 1) and \citet{Vito2022b}.}. By choosing multiple snapshots for each run, we aim to make our results more general and not depending on the specific gas/stellar morphology and AGN/star formation activity of a single snapshot.
The selected snapshots are post-processed with \code{SKIRT}\footnote{Version 8, \url{http://www.skirt.ugent.be}.} \citep{Baes:2015, Camps2016}. For the RT setup, a dust component and the radiation sources need to be specified.

Given that the processes related to the dust production and destruction are not explicitly followed in the hydrodynamic simulations adopted in this work, the dust distribution is assumed to track the one of the metals. We assume a linear scaling \citep[e.g.][]{Draine:2007} parameterized by the dust-to-metal ratio $f_{\rm d} = M_{\rm d} / M_Z$, where $M_{\rm d}$ is the dust mass and $M_Z$ is the total mass of all the metals in each gas particle. This parameter acts as a normalization factor for the overall dust content. In order to make our results less dependent on the specific choice of this parameter, we adopt two values for $f_{\rm d}$: a MW-like value, $f_{\rm d} = 0.3$, and a lower value, $f_d=0.08$, found to reproduce the observed SED of a ${z\sim 8}$ galaxy \citep{Behrens:2018}. We adopt dust optical properties of the Small Magellanic Cloud (SMC, \citealt{Weingartner:2001}). We assume gas particles hotter than $10^6$~K to be dust-free because of thermal sputtering \citep[e.g.][]{Draine:1979}. 

\begin{figure*}
    \centering
    \includegraphics[width=0.96\textwidth]{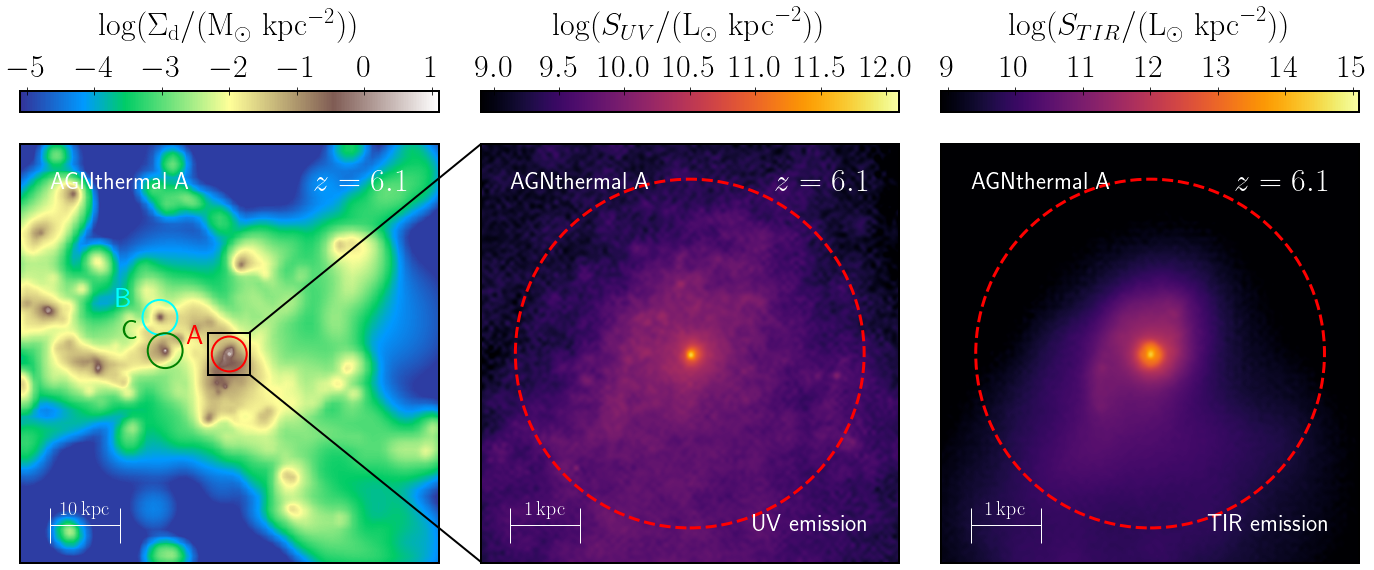}
    \hfill
    \caption{Maps of relevant properties relative to a representative snapshot, the $z=6.1$ one of the run \AGNthermal{}. \emph{Left panel:} Surface density distribution of the dust component within the computational box of $60$~kpc size. The coloured circles indicate the $2.5$~kpc region around each quasar-host with $L_{\rm bol}>10^{10}~\lsun$, which we use to compute the dust temperature PDF in Section~\ref{sec:Tdust_real} and the synthetic SEDs in Section~\ref{sec:sed_fitting}. \emph{Middle panel:} UV emission predicted by our RT calculations around source A in a smaller region of $6$~kpc size. The radius of $2.5$~kpc is shown with a dashed red line. \emph{Right panel:} Same as the middle panel, but for the TIR emission.
    \label{fig:maps_example}
    }
\end{figure*}

Dust is distributed in the computational domain in an octree grid, whose maximum number of levels of refinement for high dust density regions is chosen according to the spatial resolution of the hydrodynamic simulations. For the runs from \citetalias{paramita:2018}, we adopt $8$ levels of refinement, achieving a spatial resolution of ${\approx 230}$~pc in the most refined cells, comparable with the softening length in the hydrodynamic simulation (${\approx 210}$~pc at $z=6$); for the runs from \citetalias{Valentini:2021}, we adopt $10$ levels of refinement, with the highest resolved cells having a size of ${\approx 59}$~pc, consistent with the softening length (${\approx 87}$~pc at $z=6$). For illustration purposes, in Fig.~\ref{fig:maps_example} we show the dust surface density distribution for the snapshot at $z=6.1$ of the run \AGNthermal{}, which we consider as a representative case. Within the simulated computational box, three quasar-hosts galaxies with $L_{\rm bol}>10^{10}~\lsun$ are present, and are labelled in the figure.

We make use of the stellar synthesis models by \citet{Bruzual:2003} to implement the stellar radiation, according to the mass, age and metallicity of each stellar particle. For the black holes, we use the composite power-law AGN SED we introduced in \citetalias{DiMascia2021}, which is derived on the basis of several observational and theoretical works \citep{Shakura:1973, Fiore:1994, Richards2003AJ, Sazonov:2004, Piconcelli:2005, Gallerani:2010, Lusso:2015, Shen:2020}.
The AGN SED reads:
\begin{equation}\label{AGN_SED_eq2}
     L_\lambda = c_i \ \left(\frac{\lambda}{\mu{\rm m}}\right)^{\alpha_i} \ \left(\frac{L_{\rm bol}}{\lsun}\right) \ \lsun \ {\mum}^{-1},
\end{equation}
where $i$ labels the bands in which we decompose the spectra and the coefficients $c_i$ are determined by imposing the continuity of the function based on the slopes $\alpha_i$, as detailed in Table 2 in \citetalias{DiMascia2021}. In particular, we make use of the \emph{fiducial} AGN SED, characterized by an UV spectral slope $\alpha_{\rm UV} = -1.5$.
The SED is then normalized according to the bolometric luminosity of the AGN (see eq. \ref{eq:luminosity_bh}). The radiation field is sampled by using a grid composed of $200$ logarithmically spaced bins, covering the \emph{rest-frame} wavelength range ${[0.1-10^3]~\mum}$, with $10^6$ photon packets launched from each source per wavelength bin.

In order to isolate the contribution of the AGN to dust heating, we perform each RT run with and without AGN radiation: we refer to the first group of simulations as \AGNon{}, and to the second as \AGNoff{}. The runs \AGNoff{} and \AGNon{} share the same dust/stellar content and distribution: they differ from each other solely in the AGN contribution to the radiation field.

As an illustrative case, we show in the middle and right panel of Fig.~\ref{fig:maps_example} the UV ($0.1-0.3~\mum$) and total-infared (TIR, $8-1000~\mum$) emission in a zoomed region around source A.

\section{AGN contribution to dust heating} \label{sec:dust_temperature}

We now investigate the contribution of the AGN radiation to the dust heating, by comparing the dust temperature distribution of the \AGNon{} and \AGNoff{} runs. We first analyze the actual dust temperature in the host galaxies in Section~\ref{sec:Tdust_real}; then we discuss in Section~\ref{sec:sed_fitting} the dust temperature that would be inferred for our quasar-hosts by treating our synthetic SEDs as mock observations.

\subsection{Physical dust temperature} \label{sec:Tdust_real}

\begin{figure*}
    \centering
    \hfill
    \includegraphics[width=0.95\textwidth]{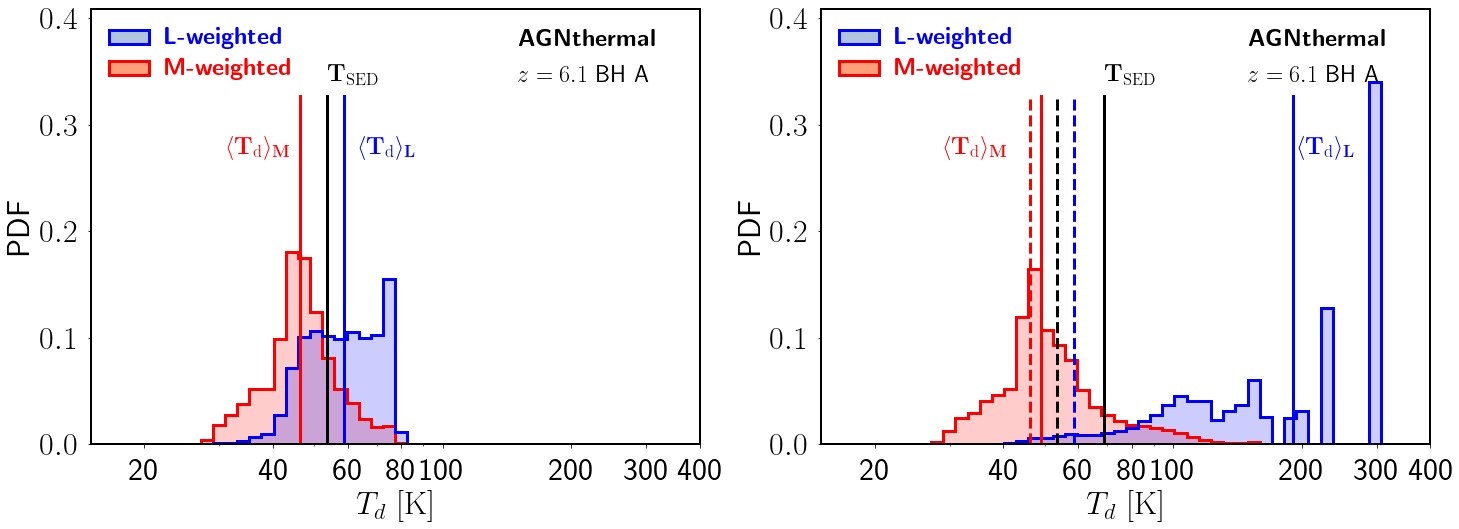}
    \caption{\emph{Left panel}: probability Distribution Function (PDF) of the mass-weighted (red) and luminosity-weighted (blue) dust temperature (see Section~\ref{sec:Tdust_real}) of the computational cells within $2.5$~kpc from a representative AGN (i.e. source A in Fig.~\ref{fig:maps_example}) in the \AGNoff{} case. The red and blue solid lines mark the median values of the mass-weighted and luminosity weighted PDFs, $T_{\rm d,M}$ and $T_{\rm d,L}$, respectively. The black solid line indicates the value of $T_{\rm SED}$ derived from the SED fitting (see Section~\ref{sec:sed_fitting}). \emph{Right panel:} same as the left panel, but for the \AGNon{} case. For comparison, we also show in this panel the values of $T_{\rm d,M}$, $T_{\rm d,L}$ and $T_{\rm SED}$ for the \AGNoff{} case with dashed lines, adopting the same color legend as in the left panel.
    \label{fig:Tdust_example}
    }
\end{figure*}

For each RT run, we select the region within $2.5$~kpc around each AGN, which encloses the AGN host. This size is comparable to the galaxy sizes in the simulations and to the dust continuum size of the majority of the quasar-hosts observed at $z\sim 6$ \citep[e.g.][]{Venemans2020ApJ}. The dust masses in the AGN hosts are in the range $3.1 \times 10^6 - 2.0 \times 10^8~\msun$, whereas the AGN bolometric luminosities are $L_{\rm bol} = 10^{10-14}~\lsun$. We compute the probability distribution function (PDF) of the dust temperatures in this region for the \AGNoff{} and \AGNon{} runs as done in \citetalias{DiMascia2021}. For each AGN, we consider the cells in the dust grid within $2.5$~kpc from the BH, and we compute the mass-weighted dust temperature PDF $\langle T_{\rm d}\rangle_M$ by weighting the dust temperature in each cell $T_{\rm d,i}$ with its dust mass $M_{\rm d,i}$. We also compute the luminosity-weighted dust temperature PDF $\langle T_{\rm d}\rangle_L$, assuming that each cell emits as a grey-body ${L_{\rm TIR,i} \propto M_{\rm d,i} T^{4+\beta_{\rm d}}_{\rm d,i}}$, where $\beta_{\rm d}$ is the dust emissivity index (see Section~\ref{sec:sed_fitting}), which for this calculation is set to $\beta_{\rm d}=2$\footnote{We verified that varying ${1.5<\beta_{\rm d}<2.5}$ does not change significantly our results.}. Finally, we compute the median values of the PDFs within the 2.5 kpc regions, $\langle T_{\rm d}\rangle_M$ and $\langle T_{\rm d}\rangle_L$, and we indicate these medians with $T_{\rm d, M}$ and $T_{\rm d, L}$, respectively. The latter is expected to be dominated by the hottest regions in the AGN proximity, whereas the former should be more representative of the bulk of the dust in the ISM. As an example, in Fig.~\ref{fig:Tdust_example}, we compare the PDFs in the \AGNoff{} and \AGNon{} cases derived for the representative quasar-host also used in Fig.~\ref{fig:maps_example}. The presence of AGN radiation shifts the dust temperature distribution toward higher values. This effect is significant for the luminosity-weighted distribution, but it is also noticeable in the mass-weighted one.

\begin{figure*}
    \centering
    \includegraphics[width=0.95\textwidth]{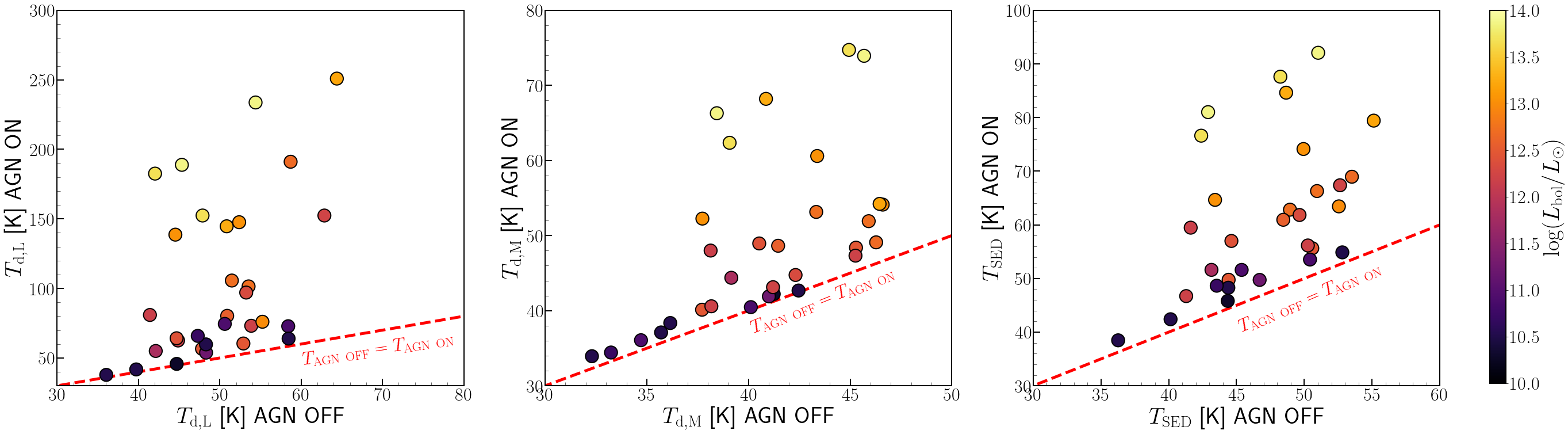}
    \hfill
    \caption{Comparison of different proxies of dust temperature distribution between the \AGNoff{} and \AGNon{} cases for all the considered AGN and snapshots. \emph{Left panel}:  median of the luminosity-weighted dust temperature $T_{\rm d, L}$; \emph{Middle}: median of the mass-weighted dust temperature $T_{\rm d, M}$; \emph{Right}: SED-derived dust temperature $T_{\rm SED}$ (see Section~\ref{sec:sed_fitting}). Each run is colour-coded according to the AGN bolometric luminosity. The dashed red lines marks the locus where temperatures in the \AGNon{} and \AGNoff{} cases are equal. 
    \label{fig:Td_scatter_2D}
    }
\end{figure*}

In order to analyze in a more systematic way the effect of the AGN radiation on the dust temperature distributions, we compare the median values of the weighted PDFs for all of the considered \AGNoff{} and \AGNon{} cases in Fig.~\ref{fig:Td_scatter_2D}. 
We find that although $T_{\rm d, L} < 70$~K in the \AGNoff{} cases, in agreement with the values usually assumed in observations to derive $M_{\rm d}$ and SFR, it increases up to $T_{\rm d, L} \approx 250$~K in the \AGNon{} runs, when AGN radiation is accounted for, showing a strong trend with AGN luminosity. This result is consistent with what found in \citet{DiMascia2021}, that AGN radiation can effectively heat dust in the ISM on $\approx 200$~pc scales from the AGN. 

A similar trend is also found for the mass-weighted dust temperature, even if less pronounced. In fact, while $T_{\rm d, M} \lesssim 50$~K in the \AGNoff{} runs, it increases up to $T_{\rm d, M} \approx 80$~K in the \AGNon{} runs. This result has important consequences: it implies that AGN can heat the bulk of the ISM dust in their host galaxies, not only in their proximity. 
The behaviour of $T_{\rm d, M}$ is also found to correlate with the AGN luminosity. In both the mass-weighted and luminosity-weighted distributions the median dust temperatures in \AGNon{} runs start to deviate (see Fig. \ref{fig:Td_scatter_2D}) from the \AGNoff{} values already at $L_{\rm bol} \approx 10^{12}~\lsun$ for most of the AGN hosts. For $L_{\rm bol} \gtrsim 10^{13}~\lsun$, $T_{\rm d, M}$ increases by $20-50\%$. This is a significant effect: since the TIR luminosity scales approximately as $L_{\rm TIR} \propto T_{\rm dust}^{4+\beta}$ (where $\beta \approx 2$ is the dust emissivity index, see Section \ref{sec:sed_fitting}), a $50\%$ boost from the AGN radiation results in a $10$-fold increase of the total IR luminosity. 
Given that most of the quasars detected at $z>6$ have bolometric luminosities larger than $10^{13}~\lsun$ \citep[e.g.][]{Yang2021ApJ...923..262Y}, our calculations suggest that AGN dominate the dust heating in the majority of these sources. 

\subsection{Dust temperature inferred from the SED} \label{sec:sed_fitting}

\begin{figure}
    \centering
    \includegraphics[width=0.45\textwidth]{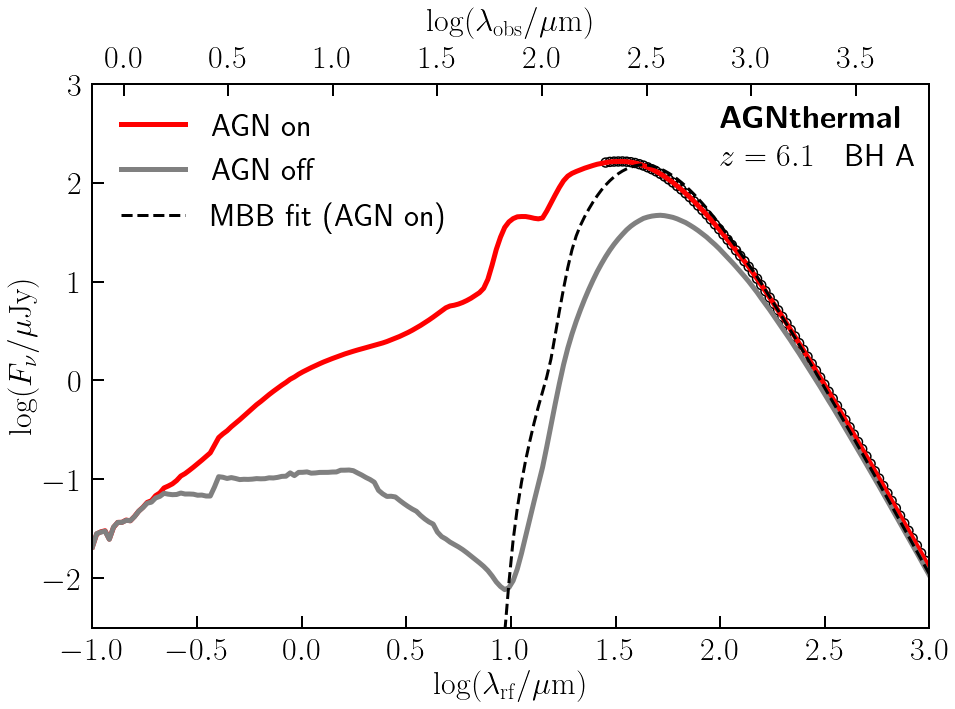}
    \hfill
    \caption{Spectral Energy Distribution (SED) emitted from the $2.5$~kpc size area selected around a representative quasar-host (source A in Fig.~\ref{fig:maps_example}). The solid red line shows the flux for the \AGNon{} RT run and the grey solid line the \AGNoff{} case. The black dashed line indicate the SED fit of the rest-frame FIR part of the SED performed according to the modified black-body function (eq.~\ref{eq:mbb_final}). The black circles indicate the points of the SED used for the fit.
    \label{fig:SED_example}
    }
\end{figure}

We derive the synthetic SED of each AGN-host by considering the radiation emitted within $2.5$~kpc from the position of the BHs, consistently with the analysis in Section \ref{sec:dust_temperature}. The SED of a representative quasar-host is shown in Fig.~\ref{fig:SED_example}. We then treat the SEDs as mock observations and we perform an SED fitting of the FIR emission.

The dust emission in the optically-thin limit can be expressed as \citep[e.g.][]{Carniani2019MNRAS}:
\begin{equation} \label{eq:mbb_final}
    S_{\nu_{\rm obs}}^{\rm obs} = \frac{1+z}{d_L^2} \ M_{\rm dust} \ \kappa_\nu \ B_\nu (T_{\rm SED}),
\end{equation}
where $S_{\nu_{\rm obs}}^{\rm obs}$ represents the observed flux at the observed frequency $\nu_{\rm obs}$ and $B_\nu (T_{\rm SED})$ is the black-body emission at the dust temperature $T_{\rm SED}$. We underline that $T_{\rm SED}$ should not be interpreted as a \quotes{true} dust temperature of the dust component in the galaxy, which has instead a complex multi-temperature distribution (see Appendix A in \citealt{Sommovigo2021MNRAS.503.4878S} for an in-depth discussion). The dust opacity $\kappa_\nu$ is usually expressed as a power-law $\kappa_0 \left(\frac{\lambda_0}{\lambda} \right)^\beta$, with the parameters $\kappa_0$, $\lambda_0$ and the dust emissivity index $\beta$ derived from theoretical models \citep[e.g.][]{Weingartner:2001, Bianchi:2007} or from observations \citep[e.g.][]{Beelen2006, Valiante2011}. We adopt the dust opacity $\kappa_\nu$ appropriate for the SMC dust model from \citet{Weingartner:2001}, the same used in the RT simulations. This leaves the dust mass $M_{\rm dust}$ and the dust temperature $T_{\rm dust}$ as the only free parameters. In most observations of $z\sim 6$ quasar-hosts only one or a few data points in the rest-frame FIR are available \citep[e.g.][]{Venemans2020ApJ}, without probing the SED peak, therefore the estimate of $T_{\rm SED}$ is very uncertain, with few notable exceptions \citep[e.g.][]{Pensabene2021}. In order to avoid this problem, we consider the FIR portion of our synthetic SEDs at $\lambda_{\rm obs}>200~\mum$ (corresponding to more than $70$ bins of the wavelength grid), which is the wavelength range typically probed by $z \sim 6$ quasars observations, and we assign a $10\%$ error to the flux at each wavelength, in order to mimic the typical experimental uncertainties \citep[e.g.][]{Venemans2020ApJ}. 

In Fig.~\ref{fig:Tdust_example} we mark the value of $T_{\rm SED}$ for the representative quasar-host shown in Fig.~\ref{fig:maps_example}. $T_{\rm SED}$ has an intermediate value between $\langle T_{\rm d}\rangle_M$ and $\langle T_{\rm d}\rangle_L$. It also tends to be more similar to the median of the mass-weighted distribution rather than to the luminosity-weighted one in the \AGNon{} cases.
In the right-most panel of Fig.~\ref{fig:Td_scatter_2D} we show how the measured values of $T_{\rm SED}$ change when including AGN radiation for all the simulated quasar-hosts. We find a significant increase of the estimated $T_{\rm SED}$ when AGN radiation is included, from $T_{\rm SED} \lesssim 60$~K in the \AGNoff{} to $T_{\rm SED} \approx 90$~K in the \AGNon{} runs. In particular, we notice that all the simulated galaxies with $T_{\rm SED} \gtrsim 60$~K are among \AGNon{} runs, suggesting that galaxies with $T_{\rm SED} \gtrsim 60$~K are likely AGN-powered. A trend with the AGN bolometric luminosity is also present, with the most luminous AGN showing the largest increase of $T_{\rm SED}$. 
This result is consistent with the behaviour of the mass-weighted and luminosity-weighted PDFs shown in Fig.~\ref{fig:Td_scatter_2D}, as $T_{\rm SED}$ partly reflects the actual dust temperature distribution in the AGN host. Therefore, the correlation found in Fig.~\ref{fig:Td_scatter_2D} is a direct consequence of the AGN heating of the ISM dust in the simulated galaxies, as seen in the behaviour of $T_{\rm d, L}$ and $T_{\rm d, M}$ (left and middle panel of Fig.~\ref{fig:Td_scatter_2D}).

\begin{figure}
    \centering
    \includegraphics[width=0.475\textwidth]{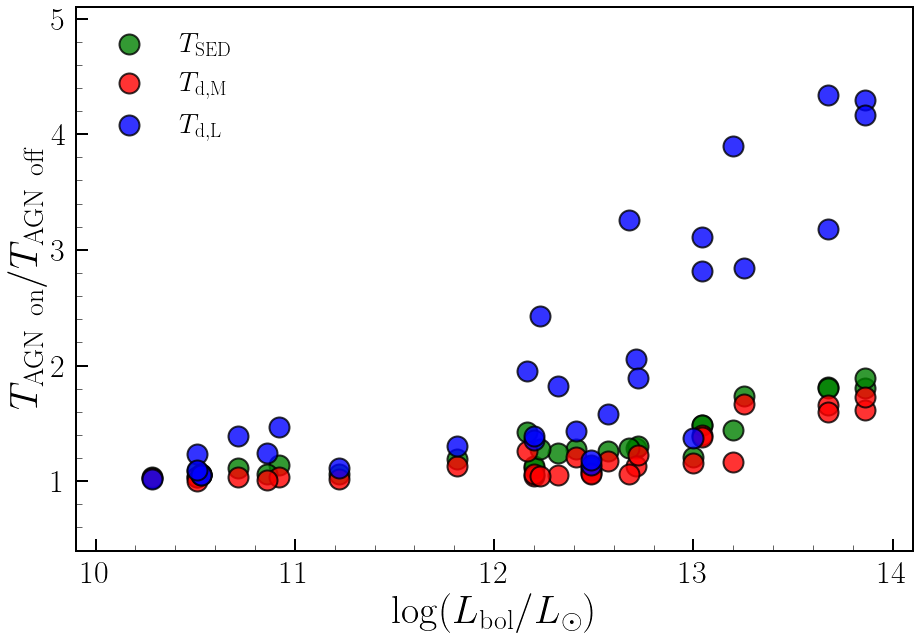}
    \caption{Ratio between the \AGNon{} and \AGNoff{} runs for the median of the luminosity-weighted temperature $T_{\rm d, L}$ (blue circles), the median of the mass-weighted temperature $T_{\rm d, M}$ (red circles), and  $T_{\rm SED}$ (green circles) as a function of the bolometric luminosity of the AGN.
    \label{fig:Td_on_to_off}
    }
\end{figure}

In Fig.~\ref{fig:Td_on_to_off} we show the ratio of the values obtained for these three indicators of the dust temperature distribution.
As expected, all the ratios increase with $L_{\rm bol}$, with the shallowest (steepest) rise for the mass-weighted (luminosity-weighted) median; $T_{\rm SED}$ is intermediate between the two. 
In particular, we note that for the brightest AGN, $L_{\rm bol} \gtrsim 10^{13}~\lsun$, the SED-derived dust temperature $T_{\rm SED}$ increases by a factor of almost $2$, similarly to the median of the mass-weighted $T_{\rm d, M}$.

\section{Star formation rate estimate from the FIR} \label{sec:SFR}

The analysis in Section~\ref{sec:dust_temperature} shows that in the simulated $z \sim 6-7$ galaxies AGN radiation can contribute significantly to dust heating, affecting the temperature of the bulk of the dust in the ISM. Given that the star formation rate estimates in high-$z$ quasar-hosts are based on the assumption that stellar radiation is the only source of dust heating in the host galaxy, we investigate the impact of the AGN contribution on the inferred SFRs values. We now take advantage of our RT simulations by using our synthetic SEDs as mock observations as in Section~\ref{sec:sed_fitting}. Then, we estimate the star formation rate in the host galaxy from the FIR luminosity, assuming that dust is heated only by stars. Finally, we compare the inferred star formation rate with the actual star formation rate in the quasar hosts, which is known from the hydrodynamic simulations.

For each simulated AGN host, we estimate the star formation rate from our synthetic SEDs as follows.
We compute the total IR luminosity $L_{\rm TIR}$ by integrating equation \ref{eq:mbb_final} over the wavelength range $8-1000~\mum$. Then, we make use of the $\SFR_{\rm FIR}-L_{\rm TIR}$ calibration in \citet{Kennicutt:2012}, based on \citet{Murphy2011ApJ}. This relation provides an estimate of the \emph{obscured} star formation rate, i.e. the stellar radiation that is absorbed by dust, \emph{assuming that dust heating is purely due to stars}. We perform this computation both for the \AGNoff{} and the \AGNon{} runs, in order to quantify the AGN contribution to the SFR estimate.

For \AGNoff{} runs we also include the \emph{unobscured} star formation rate, by adopting the $\SFR_{\rm UV}-L_{\rm FUV}$ calibration from \citet{Kennicutt:2012}\footnote{We use the results from \citet{MadauDickinson2014} -- see their Fig.~4 -- to correct the conversion factor, in order to take into account the different Initial Mass Function (IMF) adopted in the simulations \citep{Chabrier:2003} with respect to the one used in the \citet{Kennicutt:2012} calibration \citep{Kroupa2003ApJ}.}. The contribution of $\SFR_{\rm UV}$ is not accounted for in the \AGNon{} runs, consistently with observational works, since the UV emission is expected to be dominated by the quasar. We note that accounting also for $\SFR_{\rm UV}$ would increase the discrepancies between the inferred SFRs in the \AGNon{} and \AGNoff{} cases, thus reinforcing our conclusions.

\begin{figure*}
    \centering
    \includegraphics[width=0.80\textwidth]{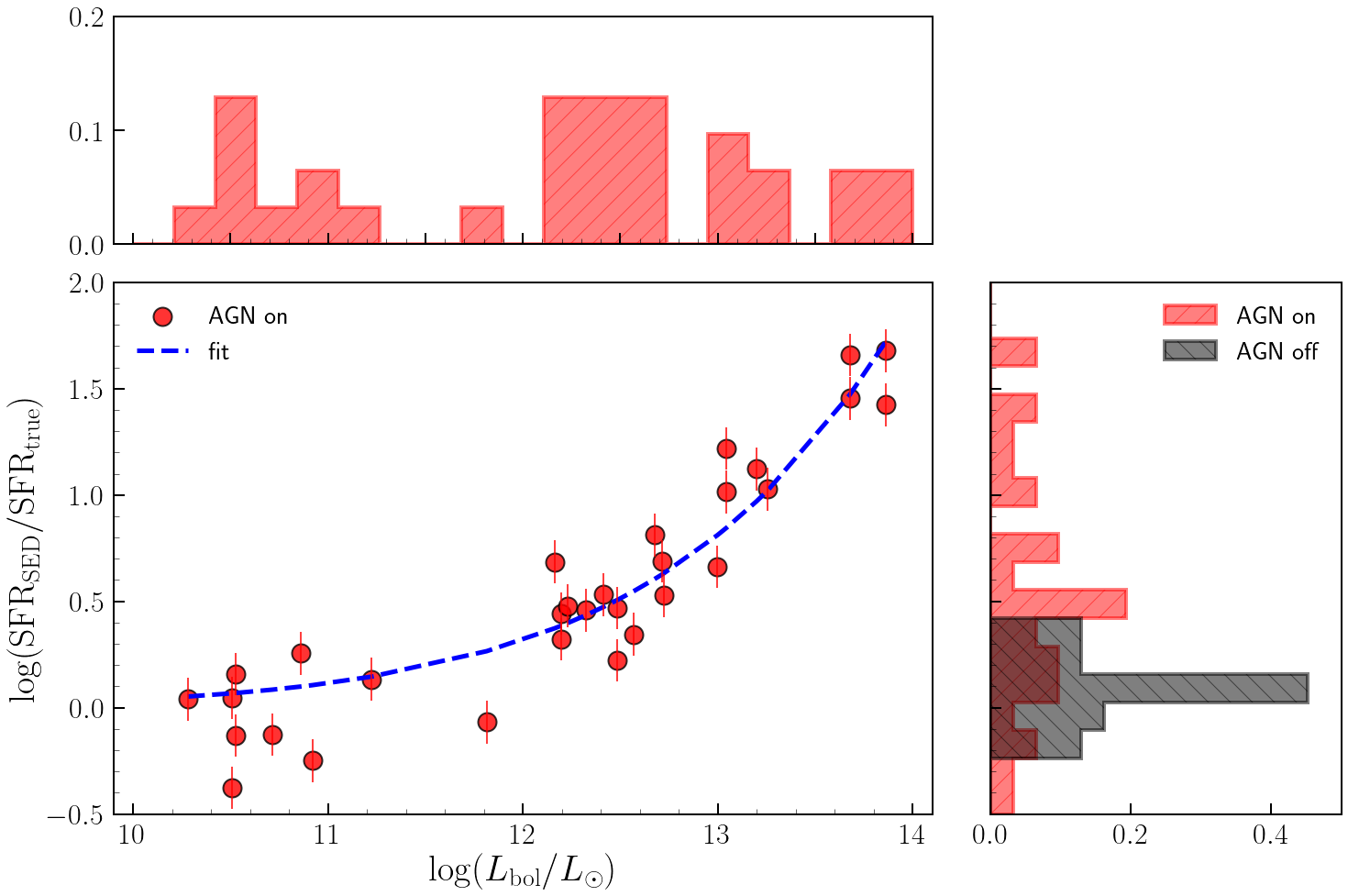}
    \caption{Ratio of the total SFR (estimated via FIR SED fitting, as explained in Section~\ref{sec:SFR}) to the true SFR in the AGN hosts as a function of the luminosity of the AGN. The blue dashed line indicates the exponential fit as in eq. \ref{eq:exp_fit}. The upper inset shows the distribution of AGN luminosities. The inset on the right shows the distribution of the ${\rm SFR}_{\rm SED}$ to ${\rm SFR}_{\rm true}$ ratio for the \AGNon{} and \AGNoff{} runs in red and black histograms respectively. We note that the distribution of the SFR estimates of the \AGNoff{} runs is consistent with the actual SFR.
    \label{fig:scatter_SFR_vs_Lbol}
    }
\end{figure*}

We compare the SFR inferred from the SED as described above with the actual SFR in the AGN hosts, which is known from the hydrodynamical simulations. The SFR is computed in the region within $2.5$~kpc from the position of each BH in order to be consistent with the synthetic SED. We consider the SFR averaged over the past $100$~Myr, which we found to provide the best agreement with the UV calibration in the \AGNoff{} runs. Fig.~\ref{fig:scatter_SFR_vs_Lbol} presents the ratio between the SED-based SFR and the true SFR as a function of  the AGN bolometric luminosity.
For the runs without AGN radiation the SFR estimated from the SED fitting is overall in agreement with the actual star formation rate in the AGN hosts (see black histogram in the right panel of Fig.~\ref{fig:scatter_SFR_vs_Lbol}). This consistency check demonstrates that our procedure correctly recovers the SFR in the host galaxy if dust heating due to AGN radiation is negligible.

In the \AGNon{} runs, the SFR inferred from the SED tends to over-estimate the actual star formation rate, with a discrepancy that becomes significant (a factor $\approx 3$) at bolometric luminosities $L_{\rm bol} = 10^{12}~\lsun$, and approximately an order of magnitude of discrepancy for most of the runs at $L_{\rm bol} \gtrsim 10^{13}~\lsun$. 
This behaviour is consistent with our findings in Section \ref{sec:dust_temperature}, in particular with the trend of $T_{\rm SED}$ between \AGNon{} and \AGNoff{} runs as a function of AGN luminosity. As discussed in Section \ref{sec:dust_temperature}, an increase by a factor of $\approx 2$ in the dust temperature results in a $\approx 10\times$ increase of the total infrared luminosity. Given that the IR luminosity is assumed to come from dust-reprocessed stellar light only, also the inferred SFR increases by the same factors.

We fit the relation $\log({\rm SFR}_{\rm SED}/{\rm SFR}_{\rm true})$ vs $\log(L_{\rm bol}/\lsun)$ with a power-law expression, finding:
\begin{equation} \label{eq:exp_fit}
    \log\left(\frac{{\rm SFR}_{\rm SED}}{{\rm SFR}_{\rm true}}\right) = 7.68 \times 10^{-14} \ \left[\log\left(\frac{L_{\rm bol}}{\lsun}\right) \right]^{11.7}.
\end{equation}
This relation quantifies the over-estimate of the star formation rate in AGN-hosts at $z \gtrsim 6$, as a function of their bolometric luminosity. It can be applied to correct the SFR in quasar-hosts with observed luminosity in the range $10^{10} \lesssim L_{\rm bol}/\lsun \lesssim 10^{14}$, and a well-constrained $T_{\rm SED}$ from the SED fitting. We assign a $0.15$~dex uncertainty to this expression based on the scatter between the same run performed with different values of the dust-to-metal ratio.

As a practical case, we apply this correction to the quasar J2348-3054, recently studied in \citet{Walter2022ApJ}. Instead of $\SFR \approx 4700~\msunyr$, we find a $\SFR \approx 370~\msunyr$, in broad agreement with the \CII-inferred $\SFR \approx 530~\msunyr$ from the relation by \citet{Herrera-Camus2018ApJ}. 

\section{Conclusions} \label{sec:conclusions}

We investigated whether stellar radiation can be considered the main source of dust heating in $z\sim 6$ quasar-hosts, and the reliability of the star formation rate (SFR) inferred from the total infrared (TIR, $8-1000~\mum$) emission in these galaxies. 
We combined cosmological hydrodynamic simulations with radiative transfer (RT) calculations, and simulated each galaxy with and without AGN radiation, in order to isolate the AGN contribution to dust heating.

We find that AGN with $L_{\rm bol} \gtrsim 10^{13}\,\lsun$ effectively heat the bulk of the dust in the interstellar medium (ISM) on galaxy scales, and not only the dust on $\approx 100$~pc from their surrounding. As a result, the actual IR emission that comes from dust-reprocessed stellar light can be significantly over-estimated.

We quantify this effect by applying the $\SFR-L_{\rm TIR}$ relation by \citet{Kennicutt:2012} to the synthetic Spectral Energy Distribution (SEDs) of the simulated quasar-hosts, and compare the results with the \quotes{true} $\SFR$ in the hydrodynamic simulations. We find that the $\SFR$ tends to be overestimated by a factor of $\approx 3$ ($\approx 30$) for $L_{\rm bol} \approx 10^{12}~\lsun$ ($L_{\rm bol} \approx 10^{13}~\lsun$, see Fig.~\ref{fig:scatter_SFR_vs_Lbol}). 
We also provide a simple relation (eq.~\ref{eq:exp_fit}) that quantifies the overestimate of the SFR in terms of the AGN luminosity.

We note that our results might be sensitive to the spatial resolution of the hydrodynamic simulations adopted in this work, the assumed models for the ISM physics, and to the numerical setup of the RT calculations. It would be valuable to repeat this analysis using simulations achieving higher spatial resolutions and a different implementation of the radiative transfer calculations.

\section*{acknowledgements}
S.C is supported by European Union’s HE ERC Starting Grant No. 101040227 - WINGS. SG acknowledges support from the ASI-INAF n. 2018-31-HH.0 grant and PRIN-MIUR 2017. AP and AF acknowledge support from the ERC Advanced Grant INTERSTELLAR H2020/740120 (PI: Ferrara). Any dissemination of results must indicate that it reflects only the author's view and that the Commission is not responsible for any use that may be made of the information it contains. MV is supported by the Alexander von Humboldt Stiftung and the Carl Friedrich von Siemens Stiftung. MV also acknowledges support from the Excellence Cluster ORIGINS, which is funded by the Deutsche Forschungsgemeinschaft (DFG, German Research Foundation) under Germany's Excellence Strategy - EXC-2094 - 390783311. We acknowledge usage of the Python programming language \citep{python2,python3}, Astropy \citep{astropy}, Cython \citep{cython}, Matplotlib \citep{matplotlib}, NumPy \citep{numpy}, \code{pynbody} \citep{pynbody}, and SciPy \citep{scipy}. We gratefully acknowledge computational resources of the Center for High Performance Computing (CHPC) at Scuola Normale Superiore, Pisa (IT). 

\section*{Data availability}
The derived data generated in this research will be shared on reasonable request to the corresponding author.

\bibliographystyle{mnras}
\bibliography{file_bibliography/ref}

\appendix

\bsp
\label{lastpage}

\end{document}

%% file: include_tex/pacchetti.tex
\usepackage[english]{babel}
\usepackage{amsmath,amssymb,bm}

\usepackage{graphicx}

\usepackage[normalem]{ulem}

\usepackage{verbatim}
\usepackage{color}
\usepackage{multirow}
\usepackage{mathtools}
\usepackage{epstopdf}

\usepackage{url}
\usepackage{xcolor}

\usepackage{gensymb}

\usepackage[usestackEOL]{stackengine}

\usepackage{multirow}

\usepackage{comment}

%% file: include_tex/journals.tex
\def\nat{Nature}

\def\mnras{MNRAS}
\def\apj{ApJ}
\def\apjl{ApJL}
\def\apjs{ApJS}
\def\aap{A\&A}

\def\araa{ARA\&A}
\def\aj{AJ}

\def\pasa{Publ. Astr. Soc. Australia}
\def\pasp{Publ. Astr. Soc. Pac.}

%% file: include_tex/definizioni.tex
\def\be{\begin{equation}} 
\def\ee{\end{equation}} 
\def\ba{\begin{eqnarray}} 
\def\ea{\end{eqnarray}}

\def\cc{\,{\rm {cm^{-3}}}} 
\def\msun{{\Msun}}

\def\HH{${\rm {H_2}}$}

\def\gsim{\lower.5ex\hbox{\gtsima}} 
\def\lsim{\lower.5ex\hbox{\ltsima}} \def\gtsima{$\; \buildrel > \over 
\sim \;$} \def\ltsima{$\; \buildrel < \over \sim \;$} \def\prosima{$\; 
\buildrel \propto \over \sim \;$} \def\gsim{\lower.5ex\hbox{\gtsima}} 
\def\lsim{\lower.5ex\hbox{\ltsima}} 
\def\simgt{\lower.5ex\hbox{\gtsima}} 
\def\simlt{\lower.5ex\hbox{\ltsima}} 
\def\simpr{\lower.5ex\hbox{\prosima}} \def\la{\lsim} \def\ga{\gsim} 
  
 \def\gtsima{$\; \buildrel > \over \sim \;$} 
\def\ltsima{$\; \buildrel < \over \sim \;$} 
\def\gsim{\lower.5ex\hbox{\gtsima}} 
\def\lsim{\lower.5ex\hbox{\ltsima}} 
\def\simgt{\lower.5ex\hbox{\gtsima}} 
\def\simlt{\lower.5ex\hbox{\ltsima}} 
\def\simpr{\lower.5ex\hbox{\prosima}} 
\def\la{\lsim} 
\def\ga{\gsim}

\def\msun{\,{\rm \Msun}}

\def\E3{{\cal E}_{\rm g}^{III}}

\def\r12{r_{1/2}} 
\def\x12{x_{1/2}} 
\def\v12{v_{1/2}}

%% file: include_tex/additional_def.tex
\def\CII{\hbox{[C~$\scriptstyle\rm II $]}}

\newcommand\code[1]{\textsc{\MakeLowercase{#1}}}
\newcommand{\quotes}[1]{``#1''}

\def\nh2{n_{\rm H2}}
\def\fh2{f_{\rm H2}}

\def\angstrom{\textrm{A\kern -1.3ex\raisebox{0.6ex}{$^\circ$}}}

\def\mum{\mu{\rm m}}
\def\msun{{\rm M}_{\odot}}

\def\lsun{{\rm L}_{\odot}}

\def\cc{{\rm cm}^{-3}}

\def\msunyr{\msun\,{\rm yr}^{-1}}

\def\SFR{{\rm SFR}}

\def\AGNsphere{\emph{AGNsphere}}
\def\AGNcone{\emph{AGNcone}}
\def\AGNthermal{\emph{AGNthermal}}

\def\AGNon{\emph{AGN on}}
\def\AGNoff{\emph{AGN off}}

%% file: include_tex/colors.tex
\makeatletter
\def\@hex@@Hex#1%
 {\if a#1A\else \if b#1B\else \if c#1C\else \if d#1D\else
  \if e#1E\else \if f#1F\else #1\fi\fi\fi\fi\fi\fi \@hex@Hex}
\makeatother

\definecolor{apcolor}{HTML}{008000}
\definecolor{vdcolor}{HTML}{ff0f00}
\definecolor{afcolor}{HTML}{b3443c}
\definecolor{sgcolor}{HTML}{b3003b}
\definecolor{sccolor}{HTML}{ffbe33}
\definecolor{fvcolor}{HTML}{ce33ff}
\definecolor{pbcolor}{HTML}{333300}
\definecolor{fdcolor}{HTML}{ff6600}